\documentstyle[12pt,aaspp4]{article}

\begin{document}


\title{The Light-cone Effect on the Shapiro Time Delay} 

\author{Hideki Asada}
\affil{ 
Faculty of Science and Technology, 
Hirosaki University, Hirosaki 036-8561, Japan
} 
\authoremail{asada@phys.hirosaki-u.ac.jp}

\begin{abstract}
We investigate the light-cone effect on the Shapiro time delay. 
The extra time delay caused by Jupiter on the 8th of September 2002 
can be measured by advanced VLBI (very long baseline interferometry). 
Our expression for the delay is in complete agreement with 
that of Kopeikin (2001), in which he argued that the excess time delay 
was due to the propagation of gravity. The present letter, however, 
shows that the excess comes from nothing but the propagation of light, 
namely the light-cone effect. 
To make a robust confirmation of general relativity by the coming 
Jupiter event, it is important to take account of the light-cone effect 
on the Shapiro time delay. 
\end{abstract}

\section{Introduction}
The Shapiro time delay plays an important role in experimental 
verification of Einstein's theory of general relativity. 
In fact, VLBI (very long baseline interferometry) confirms 
the validity of general relativity. 
The accuracy will be achieved within a few picoseconds (ps), 
namely about 10 microarcseconds $(\mu as)$, 
for instance by VERA (VLBI Exploration of Radio Astrometry; 
see Homma, Kawaguchi and Sasao 2000). 
This is why the correction to the Shapiro time delay has been
intensively investigated (Kopeikin and Sch\"afer 1999, Kopeikin 2001). 
Kopeikin found that the excess time delay caused by Jupiter 
can be measured on the 8th of September 2002. He also argued 
that the excess was due to the propagation of gravity, 
which could be tested through the observation. 
The present letter, however, shows that it comes from 
the propagation of light but not gravity. 

The primary reason against his conclusion is based on the
post-Newtonian approximation of general relativity: 
In the approximation, we perform expansions in the inverse of 
light velocity $c$, by considering all quantities are 
perturbations around the Newtonian parts.  
The deflection of light, the perihelion shift of Mercury and 
the time delay occur at the first post-Newtonian order $O(c^{-2})$. 
Actually, using these three effects, the classical tests 
have been done to confirm the validity of general relativity 
(For a thorough review, see Will 1993). 
The propagation of gravity appears at $O(c^{-4})$ as 
gravitational waves (For instance, Thorne 1980, Will 1993). 
The effect of the radiation reaction of quadrupole gravitational 
waves (Peters and Mathews 1963, Thorne 1980, Will 1993) 
has been confirmed through the observation of decaying orbital 
period of Hulse-Taylor binary pulsar (Taylor 1994). 
On the other hand, the Kopeikin's excess time delay 
for the standard Shapiro delay at $O(c^{-2})$ is $O(v_J/c^3)$, 
where $v_J$ is the velocity of Jupiter. The order of the excess 
is lower than that of the propagation of gravity, so that 
the excess cannot be caused by the gravity propagation. 
What is the origin of the excess ? The purpose of the present letter
is to answer it.

\section{Shapiro delay in retarded time}
Since a massive body produces gravitational fields as a curved 
spacetime, a light signal will take a longer time to traverse 
a given spatial distance than it would if Newtonian theory were valid. 
In deriving the Shapiro time delay, the Einstein equation for 
the gravitational field and the null geodesics for the light ray 
are solved up to $O(c^{-2})$. In particular, the Einstein equation 
is reduced to Poisson-type equations, so that the propagation 
of gravity is not incorporated. The Shapiro delay for 
a light signal from an emitter to an observer is obtained 
in a logarithmic form (Shapiro 1964, Will 1993). 

Let us consider a baseline denoted by $\mbox{\boldmath $B$}$; 
at $t_1$ and $t_2$, the light signals from a quasar reach 
the first and second stations which locate at 
$\mbox{\boldmath $x$}_1(t)$ and $\mbox{\boldmath $x$}_2(t)$, 
respectively, so that we can define the baseline as 
the spatial interval between the simultaneous events 
$\mbox{\boldmath $B$}=\mbox{\boldmath $x$}_1(t_1)
-\mbox{\boldmath $x$}_2(t_1)$. 
Each station is denoted by $i$ later. 
Since the Shapiro delay is a consequence of integration of 
the null geodesics on the light cones, it is convenient and 
crucial to use $s_1$ and $s_2$, retarded time which is constant 
on each light cone emanating from events 
$(t_1, \mbox{\boldmath $x$}_1(t_1))$ 
and $(t_2, \mbox{\boldmath $x$}_2(t_2))$, 
so that we can have 
$\mbox{\boldmath $x$}_i(t_i)=\mbox{\boldmath $x$}_i(s_i)$ 
for $i=1, 2$. 
Hence, the difference of the Shapiro delay between the baseline 
is expressed as \begin{equation}
\Delta(t_1,t_2)=\frac{2GM}{c^3}\ln
\frac{R_{1J}+\mbox{\boldmath $K$}\cdot\mbox{\boldmath $R$}_{1J}}
{R_{2J}+\mbox{\boldmath $K$}\cdot\mbox{\boldmath $R$}_{2J}} , 
\label{eq:shapiro}
\end{equation}
where the unit vector from the Earth to the emitter of the light 
is denoted by $\mbox{\boldmath $K$}$, the position of Jupiter 
by $\mbox{\boldmath $x$}_J(t)$ and we defined 
$\mbox{\boldmath $R$}_{iJ}=\mbox{\boldmath $x$}_i(s_i)
-\mbox{\boldmath $x$}_J(s_i)$ and 
$R_{iJ}=|\mbox{\boldmath $R$}_{iJ}|$ on each light cone 
labeled by $i=1, 2$. 

Since the speed of Jupiter $v_J$ is much {\it smaller} than $c$, 
we find 
\begin{eqnarray}
\mbox{\boldmath $R$}_{iJ}
&=&\mbox{\boldmath $r$}_{iJ}
+\frac{\mbox{\boldmath $v$}_J}{c}r_{iJ}+O(c^{-2}) , \\
R_{iJ}&=&r_{iJ}+\frac{\mbox{\boldmath $R$}_{iJ}
\cdot\mbox{\boldmath $v$}_J}{c}+O(c^{-2}) , 
\end{eqnarray}
where we used 
$\mbox{\boldmath $x$}_i(s_i)=\mbox{\boldmath $x$}_i(t_i)$ 
and denoted the spatial displacement vector between 
the simultaneous events by 
\begin{equation}
\mbox{\boldmath $r$}_{iJ}=\mbox{\boldmath $x$}_i(t_i)
-\mbox{\boldmath $x$}_J(t_i), 
\end{equation}
and the interval by $r_{iJ}=|\mbox{\boldmath $r$}_{iJ}|$. 
Furthermore, since $B$ is much {\it shorter} than 
$R \sim R_{iJ}$, we obtain 
\begin{equation}
r_{2J}-r_{1J}=\mbox{\boldmath $N$}_{1J}\cdot\mbox{\boldmath $B$}
+O\left(\frac{B^2}{R}\right) , 
\end{equation}
where we defined 
$\mbox{\boldmath $N$}_{1J}=\mbox{\boldmath $r$}_{1J}/r_{1J}$. 
Hence, Eq. $(\ref{eq:shapiro})$ becomes 
\begin{eqnarray}
\Delta(t_1,t_2)=\frac{2GM}{c^3}&\Bigl(&
\ln
\frac{r_{1J}+\mbox{\boldmath $K$}\cdot\mbox{\boldmath $r$}_{1J}}
{r_{2J}+\mbox{\boldmath $K$}\cdot\mbox{\boldmath $r$}_{2J}} 
\nonumber\\
&&-\frac{\mbox{\boldmath $B$}\cdot\mbox{\boldmath $v$}_J
+(\mbox{\boldmath $N$}_{1J}\cdot\mbox{\boldmath $B$})
(\mbox{\boldmath $K$}\cdot\mbox{\boldmath $v$}_J)}
{c(r_{1J}+\mbox{\boldmath $K$}\cdot\mbox{\boldmath $r$}_{1J})}
+O(c^{-2}) \Bigr) . 
\label{eq:shapiro2}
\end{eqnarray}

We denote by $\theta$ a {\it small} angle between the source 
and the Jupiter seen at the first station. 
We obtain 
\begin{eqnarray}
\mbox{\boldmath $N$}_{1J}&=&-\mbox{\boldmath $K$}\cos\theta
+\mbox{\boldmath $n$}\sin\theta \nonumber\\
&=&-\left(1-\frac{\theta^2}{2}\right)\mbox{\boldmath $K$}
+\theta\mbox{\boldmath $n$}+O(\theta^3) , 
\end{eqnarray}
where $\mbox{\boldmath $n$}$ is a unit normal vector from the Jupiter 
to the light ray. 
Using this relation, we find 
\begin{eqnarray}
r_{1J}+\mbox{\boldmath $K$}\cdot\mbox{\boldmath $r$}_{1J}
&=&\frac{\theta^2 r_{1J}}{2}+O(\theta^4) , \\
\frac{r_{1J}+\mbox{\boldmath $K$}\cdot\mbox{\boldmath $r$}_{1J}}
{r_{2J}+\mbox{\boldmath $K$}\cdot\mbox{\boldmath $r$}_{2J}}
&=&1-\frac{2\mbox{\boldmath $n$}\cdot\mbox{\boldmath $B$}}
{r_{1J}\theta}+O\left(\frac{B^2}{r^2}\right) , 
\end{eqnarray}
where we introduced $r \sim r_{1J} \sim r_{2J}$. 
Using these approximations, Eq. $(\ref{eq:shapiro2})$ is rewritten as 
\begin{eqnarray}
\Delta(t_1,t_2)=-\frac{4GM}{c^3}&\Bigl(&
\frac{\mbox{\boldmath $n$}\cdot\mbox{\boldmath $B$}}
{r_{1J}\theta} \nonumber\\
&&+\frac{\mbox{\boldmath $B$}\cdot\mbox{\boldmath $v$}_J
-(\mbox{\boldmath $K$}\cdot\mbox{\boldmath $B$})
(\mbox{\boldmath $K$}\cdot\mbox{\boldmath $v$}_J)}
{c r_{1J}\theta^2}
+O(c^{-2}, c^{-1}B^2r^{-2}) \Bigr) , 
\label{eq:shapiro3}
\end{eqnarray}
which is in complete agreement with Eq. (12) of Kopeikin (2001). 
In order to put some experimental constraints on theories of gravity, 
it is convenient to introduce PPN parameters (Will 1993). 
In this case, the numerical coefficients in front of R. H. S. of
Eqs. ($\ref{eq:shapiro}$) and ($\ref{eq:shapiro3}$), 2 and 4, should 
be replaced by  $1+\gamma$ and $2(1+\gamma)$, respectively.  

For the event caused by Jupiter in September 2002, the first and 
second terms of Eq. ($\ref{eq:shapiro3}$) become about 100 and 10 ps, 
respectively (Kopeikin 2001). 
Furthermore, it is noteworthy that they are proportional 
to $\theta^{-1}$ and $\theta^{-2}$ respectively, so that we can 
separate them if observation is made during some interval. 
In deriving Eq. ($\ref{eq:shapiro3}$), we take account of 
the propagation only of light but not of gravity, since 
gravity propagation appears at $O(c^{-4})$. 
Hence, it turns out that the excess time delay given by 
Eq. $(\ref{eq:shapiro3})$ is due to nothing but the light-cone effect. 

Before closing this section, let us make two comments: 
First, Kopeikin (2001) is based on a general formulation developed by 
Kopeikin and Sch\"afer (1999), in which both of the Einstein equation 
and the geodesics are treated carefully as propagation equations. 
As a result, effects of propagation of both gravity and light 
may be included. Therefore, it is quite difficult to distinguish 
the propagation of gravity from that of light in their resultant 
equations, and clarify the origin of the excess time delay. 
On the other hand, our Eq. $(\ref{eq:shapiro})$, the standard formula 
for the Shapiro delay, can be derived from the post-Newtonian 
metric which is obtained by solving the Poisson-type equation 
for gravitational fields (Will 1993). 
Consequently, it is obvious in our approach that the hyperbolicity 
of the Einstein equation like a wave equation plays no role 
in the excess time delay. 
It follows that our Eq. ($\ref{eq:shapiro}$) coincides with 
Eq. (6) in Kopeikin (2001) by neglecting a factor 
$(1+\mbox{\boldmath $K$}\cdot\mbox{\boldmath $v$}_J/c)$, 
since $v_J/c$ is of the order of $10^{-4}$, 
which is too small to detect. 
On the other hand, the excess term in Eq. ($\ref{eq:shapiro3}$), 
which originates from the light-cone effect, causes a sizable
correction of the order of $v_J/c\theta$, say $10^{-1}$ 
for a small angle $\theta\sim 10^{-3}$. 

Secondly, it is worthwhile to take another point of view; 
Eq. $(\ref{eq:shapiro})$ can be derived also by using 
the Schwarzschild metric (Shapiro 1964), a static solution of 
the Einstein equation, if we keep only the linear term of the mass 
and consider that the position of observers depends on light cones 
labeled by retarded time. In this case, the velocity 
of the Earth will appear instead of that of the Jupiter. 
Hence, both investigations utilizing the post-Newtonian approximation 
fixing the observer and the Schwarzschild metric fixing the Jupiter 
imply that $\mbox{\boldmath $v$}_J$ should be considered as 
the {\it relative} velocity of the Jupiter to the Earth. 
In fact, since $\mbox{\boldmath $R$}_{iJ}$ and $\mbox{\boldmath $K$}$ 
in Eq. ($\ref{eq:shapiro}$) are relative vectors, vectors 
in Eq. $(\ref{eq:shapiro3})$ are too. This is not mentioned 
in Kopeikin (2001).

\section{Conclusion}
We have examined the light-cone effect on the Shapiro time delay. 
Our expression for the extra time delay is in complete agreement 
with that of Kopeikin (2001), in which he argued that the excess 
time delay was due to the propagation of gravity. 
However, we have taken account of the light cone effect but not the 
propagation of gravity. Hence, it has been clearly shown that 
the excess comes from nothing but the propagation of light. 
To make a robust confirmation of general relativity by the Jupiter 
event on the 8th of September 2002, it is crucial to take account of 
the light-cone effect on the Shapiro time delay.

\acknowledgements
The author would like to thank M. Homma for pointing out 
the observational impact of the work by Kopeikin (2001). 
He would be also grateful to M. Kasai for useful conversation. 
This work was supported by a Japanese Grant-in-Aid 
for Scientific Research from the Ministry of Education, 
No. 13740137 and the Sumitomo Foundation.

\end{document}